\documentclass[aps,twocolumn,groupedaddress]{revtex4}
\usepackage{amssymb,amsmath}
\usepackage{graphicx,array}
\usepackage{dcolumn}
\usepackage{bm}

\begin{document}
\title{Lattice effects on the formation of oxygen vacancies in perovskite thin films}

\author{Claudio Cazorla}
\affiliation{School of Materials Science and Engineering, UNSW Australia, Sydney NSW 2052, Australia \\
	     Integrated Materials Design Centre, UNSW Australia, Sydney NSW 2052, Australia}

\begin{abstract}
We use first-principles methods to investigate the effects of collective lattice excitations
on the formation of oxygen vacancies in perovskite thin films. We find that phonons play a 
crucial role on the strain-mediated control of defect chemistry at finite temperatures. In 
particular, zero-temperature oxygen vacancy formation trends deduced as a function of 
epitaxial strain can be fully reversed near room temperature. Our first-principles calculations 
evidence a direct link between the lattice contribution to the oxygen vacancy free energy and the 
volume expansion that the system undergoes when is chemically reduced: The larger the resulting 
volume expansion, the more favorable thermal excitations are to point defect formation. However, 
the interplay between the vibrational vacancy entropy, or equivalently, chemical expansion, and 
epitaxial strain is difficult to generalise as this can be strongly influenced by underlying 
structural and magnetic transitions. In addition, we find that vacancy ordering can be largely 
hindered by the thermal lattice excitations. 
\end{abstract}

\maketitle

\begin{figure}[t]
\centerline{
\includegraphics[width=0.95\linewidth]{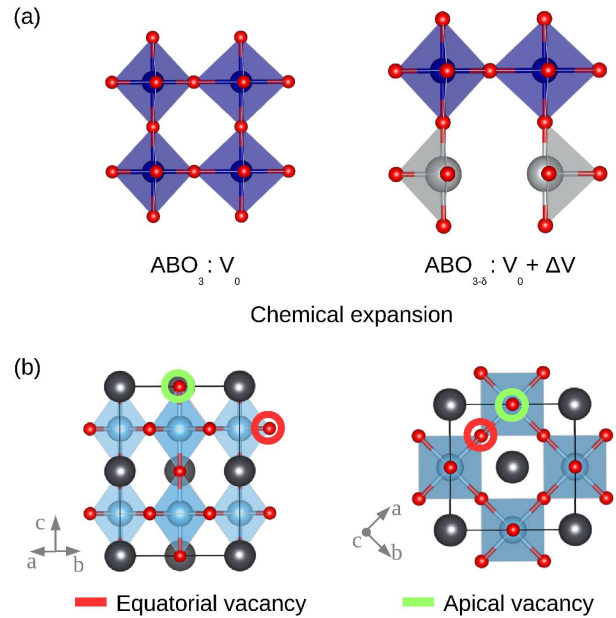}}
\vspace{0.25cm}
\caption{(Color online)~Sketch of the chemical expansion occurring in ABO$_{3}$ perovskites due to  
         oxygen reduction~(a). (b)~The $20$-atom $\sqrt{2} \times \sqrt{2} \times 2$ 
         simulation cell used in our zero-temperature energy calculations; red, 
         blue, and black spheres represent O, B, and A atoms, respectively. 
         Equatorial and apical oxygen positions are noted in the figure.}
\label{fig1}
\end{figure}

Point defects can affect considerably the structural, magnetic, and transport 
properties of perovskite-structure materials with chemical formula ABO$_{3}$. Oxygen 
vacancies (${\rm V_{O}}$), for example, enable ionic conductivity in perovskite-based 
solid solutions to be used for electrochemical applications such as solid oxide fuel and 
electrolysis cells~\cite{shao04,jeen13,tsvetkov16}. Likewise, ${\rm V_{O}}$ can significantly 
distort the equilibrium arrangement of atoms and hence modify the superexchange interactions 
between neighboring magnetic ions~\cite{taguchi79,goodenough04,cao09}. The presence of point 
defects in perovskite oxides also is known to induce an increase of volume, the so-called 
``chemical expansion''~\cite{adler01,ullmann01,marrocchelli15,aidhy15}; such an effect is 
caused by atomic underbonding due to the extra electrons provided by the oxygen vacancies, 
which are located in nonbonding transition-metal orbitals (Fig.~\ref{fig1}a). 
Engineering of defect properties in perovskite oxides, therefore, emerges as a likely avenue 
for the design of new materials with tailored functionality.  
  
Recently, it has been experimentally demonstrated that strain engineering can be used to 
tune the content of oxygen in some perovskite oxides~\cite{jeen13,hu15,petrie16,agrawal16,
hu16}. For example, in SrCoO$_{3-\delta}$ thin films a moderate epitaxial strain of about 
$+2$\% produces a $\sim 30$\% reduction in the oxygen activation energy barrier, which makes 
it possible to stabilise oxygen-deficient samples at annealing temperatures close to ambient 
conditions~\cite{petrie16}. Also, in multiferroic SrMnO$_{3}$ thin films the formation energy 
of oxygen vacancies is decreased by $0.25$~eV at a epitaxial strain of $+3.8$\%~\cite{agrawal16}.  

First-principles computational methods have been used to unveil the atomistic mechanisms 
behind strain-mediated ${\rm V_{O}}$ formation at zero temperature (i.e., neglecting 
possible thermal effects)~\cite{aschauer13,aschauer15,tahini16}. Interestingly, Aschauer 
\emph{et al.}~\cite{aschauer13} have shown that the formation energy of oxygen vacancies 
in CaMnO$_{3}$ thin films is strongly favored by tensile (positive) epitaxial strain. This 
finding has been rationalised in terms of general electrostatic arguments based on the fact 
that the accompanying electron-electron repulsion is effectively reduced along elongated 
Mn-Mn distances. Nevertheless, experiments performed in CaMnO$_{3}$ thin films covering a 
wide range of epitaxial states have not evidenced any preference for ${\rm V_{O}}$ formation 
at tensile conditions~\cite{flint14}. Further, recent measurements in SrCoO$_{3-\delta}$ thin 
films by Hu \emph{et al.}~\cite{hu16} have revealed a nonmonotonic saw-tooth dependence of 
the critical reduction temperature on epitaxial strain with a marked peak at moderate tensile 
stresses, which suggests the presence of ${\rm V_{O}}$ contributions other than purely electrostatic.     

In this Letter, we employ first-principles methods based on density functional theory (DFT)
to quantify the effects of thermal lattice excitations on the formation energy of ${\rm V_{O}}$
in perovskite thin films. We select SrCoO$_{3-\delta}$ (SCO) and 
La$_{0.5}$Sr$_{0.5}$Mn$_{0.5}$Co$_{0.5}$O$_{3-\delta}$ (LSMCO) as the reference systems in which 
to perform our calculations because these materials exhibit one of the highest oxygen-deficient 
stoichiometries observed to date at ambient conditions in a simple perovskite~\cite{takeda72,jeen13b,aguadero11}. 
Our computational results show that (1)~thermal lattice excitations play a crucial role on the 
strain-dependence of oxygen vacancy formation, as they can fully reverse the energy trends deduced 
at zero temperature; (2)~the larger the chemical expansion experienced by the system the more 
favorable vibrational entropy contributions are to ${\rm V_{O}}$ formation; and (3)~the 
vibrational vacancy entropy depends on the defect symmetry, that is, whether is equatorial or 
apical (Fig.~\ref{fig1}b), which in some systems may oppose to vacancy ordering at medium and
high temperatures. 

\begin{figure}[t]
\centerline{
\includegraphics[width=0.90\linewidth]{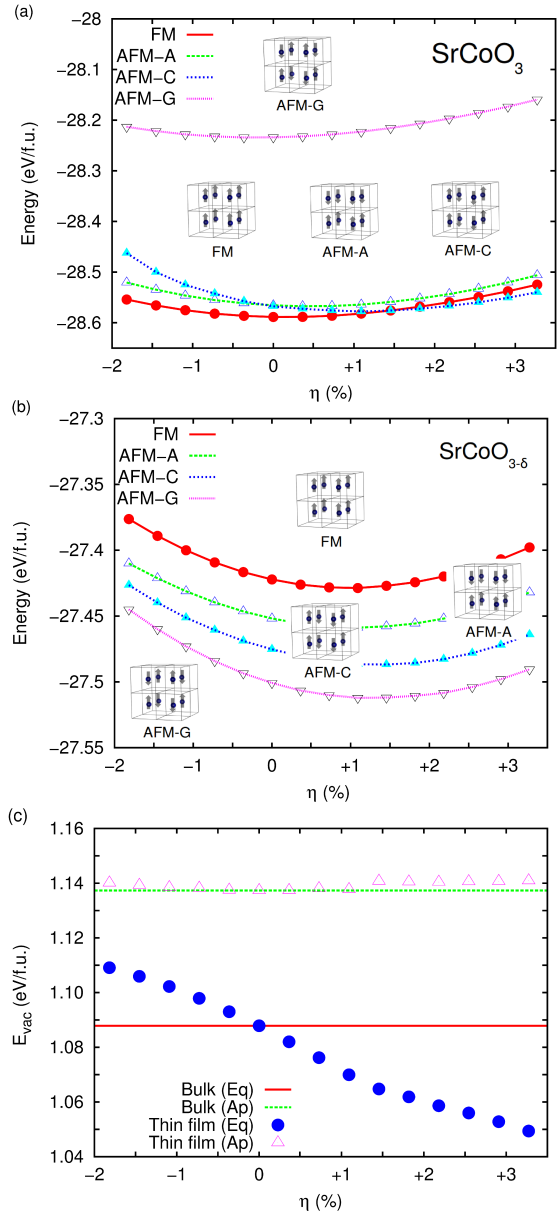}}
\caption{(Color online)~Zero-temperature calculations done in perfect and
         nonstoichiometric ($\delta = 0.25$) SCO thin films. Total energy 
         results expressed as a function of magnetic spin ordering, epitaxial 
         strain, and oxygen content are shown in (a) and (b). The    
         zero-temperature vacancy formation energy is enclosed in (c). 
         Both equatorial (Eq) and apical (Ap) oxygen vacancy positions 
         are considered and analogous results obtained in bulk are shown 
         for comparison.}
\label{fig2}
\end{figure}

\begin{figure}[t]
\centerline{
\includegraphics[width=1.0\linewidth]{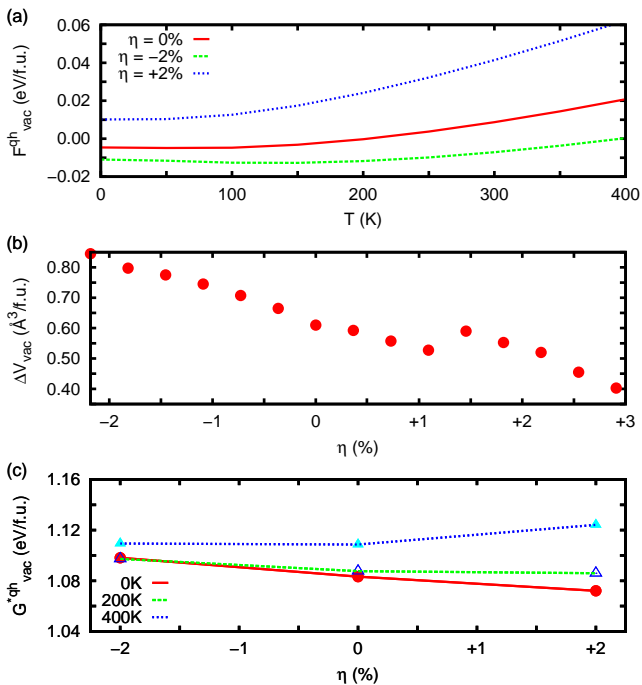}}
\caption{(Color online)~Quasi-harmonic free energy calculations performed in perfect and
         nonstoichiometric ($\delta = 0.25$, equatorial vacancies) 
         SCO thin films. Vibrational and thermodynamically shifted Gibbs 
         vacancy free energy results expressed as a function of temperature 
         and epitaxial strain are enclosed in (a) and (c), respectively. The 
         accompanying chemical expansion, $\Delta V_{\rm vac} = V_{\rm nonstoi} - V_{\rm perf}$, 
         expressed as a function of epitaxial strain is shown in (b).}
\label{fig3}
\end{figure}

We compute the quasi-harmonic Gibbs free energy associated to the formation of a 
neutrally-charged oxygen vacancy, $G^{\rm qh}_{\rm vac}$, as a function of epitaxial 
strain, $\eta \equiv \left(a - a_{0}\right) / a_{0}$ (where $a_{0}$ represents the 
equilibrium in-plane lattice parameter) and temperature, $T$. This quantity can be 
expressed as~\cite{hu16}: 
\begin{equation}
G^{\rm qh}_{\rm vac}(\eta, T) = E_{\rm vac}(\eta) + F^{\rm qh}_{\rm vac}(\eta, T) + \mu_{\rm O}(T)~, 
\label{eq1}
\end{equation}
where subscript ``vac'' indicates the quantity difference between the defective and perfect 
systems (in SCO, for instance, $A_{\rm vac} \equiv A_{\rm SrCoO_{3-\delta}} - A_{\rm SrCoO_{3}}$), 
$E_{\rm vac}$ accounts for the static contributions to the free energy ($T = 0$), $F^{\rm qh}_{\rm vac}$ 
the vibrational contributions to the free energy ($T \neq 0$), and $\mu_{\rm O}$ is the chemical 
potential of the free oxygen atom. The vibrational energy of the defective and perfect systems 
are computed with the formula~\cite{cazorla13}:  
\begin{equation}
 F^{\rm qh}(\eta, T) = \frac{1}{N_{q}}~k_{\rm B} T \sum_{{\bf
    q}s}\ln\left[ 2\sinh \left( \frac{\hbar\omega_{{\bf
        q}s}(\eta)}{2k_{\rm B}T} \right) \right]~,  
\label{eq2}
\end{equation}
where $N_{q}$ is the total number of wave vectors used for integration within the Brillouin zone 
(BZ), and the dependence of the phonon frequencies, $\omega_{\boldsymbol{q}s}$, on epitaxial strain 
is explicitly noted.  

We use the generalised gradient gradient approximation to DFT due to Perdew, Burke, and 
Ernzerhof~\cite{pbe} as implemented in the VASP package~\cite{vasp}. A ``Hubbard-U''~\cite{hubbard} 
scheme is employed for a better treatment of transition-metal $d$ electrons, and the ionic cores 
are represented with the ``projector augmented'' method~\cite{paw}. Wave functions are represented 
in a plane-wave basis truncated at $600$~eV. For the determination of equilibrium geometries we use 
a $20$-atoms simulation cell that allows to reproduce the usual ferroelectric and antiferrodistortive 
distortions in perovskite oxides~\cite{cazorla15} (Fig.~\ref{fig1}b); defective configurations are
generated by removing one oxygen atom either in an equatorial (Eq) or apical (Ap) position (case 
$\delta = 0.25$). For our phonon frequency calculations we employ the so-called ``direct approach'' as 
implemented in the PHON package~\cite{phon}. Additional details of our $E_{\rm vac}$ and 
$F^{\rm qh}_{\rm vac}$ calculations can be found in the Supplemental Material~\cite{supp} and 
Refs.~\cite{cazorla16,cazorla09,cazorla15b}. 

It is well known that estimation of $\mu_{\rm O}$ with DFT+U methods contains large errors~\cite{jones89,wang06}.
Notwithstanding, as the oxygen chemical potential does not depend on epitaxial strain, we can safely 
base our following analysis on the results obtained for the thermodynamically shifted Gibbs free energy: 
\begin{equation}
G^{* \rm qh}_{\rm vac}(\eta, T) = G^{\rm qh}_{\rm vac}(\eta, T) - \mu_{\rm O}(T)~.
\label{eq3}
\end{equation}
Namely, rather than trying to adopt experimental values for $\mu_{\rm O}$ and applying empirical
corrections to the calculated $G^{\rm qh}_{\rm vac}$, we have arbitrarily set the oxygen gas chemical 
potential to zero. Meanwhile, we have estimated the size of anharmonic corrections to $F^{\rm qh}_{\rm vac}$ 
by using thermodynamic integration techniques from harmonic reference systems (Supplemental 
Material~\cite{supp} and Refs.~\cite{cazorla09b,cazorla12}); our results show that although 
anharmonic corrections affect considerably the vibrational vacancy entropy (i.e., by $\sim 
0.01$~eV per formula unit), the main conclusions deduced from $G^{* \rm qh}_{\rm vac}$ are robust.

{\bf SrCoO$_{3-\delta}$ thin films}--According to our DFT calculations, the ground state of stoichiometric 
SCO is a tetragonal $P4/mbm$ phase with an equilibrium lattice parameter of $a_{0} = 3.89$~\AA~ and 
ferromagnetic (FM) spin ordering~\cite{cazorla16}. At $\eta \sim +2$\%, the system undergoes a 
magnetic phase transition to an antiferromagnetic state displaying C-type spin ordering (i.e., spins in the plane 
parallel to the substrate align antiparallel, whereas spins in planes perpendicular to the substrate align 
parallel, Fig.~\ref{fig2}a). In the presence of oxygen vacancies, the magnetic properties of SCO change noticeably: 
the ground state becomes antiferromagnetic of G-type (i.e., spins align antiparallel both in the substrate plane 
and perpendicular to it, Fig.~\ref{fig2}b) either at moderately tensile or compressive (negative) strains.  
With regard to $E_{\rm vac}$, we find that (at $T = 0$) equatorial vacancies can be created more easily than
apical vacancies [i.e., $E_{\rm vac}({\rm Ap})-E_{\rm vac}({\rm Eq}) \sim 10$~meV/f.u.] and that their formation 
energy decreases almost linearly with increasing strain (Fig.~\ref{fig2}c). For instance, at $\eta = +2$\% 
the (Eq) vacancy formation energy is reduced by $30$ and $52$~meV/f.u. as compared to the unstrained and 
$-2$\% cases, respectively. We note that these results are consistent with those reported by Tahini \emph{et al.} 
in analogous systems~\cite{tahini16}, and by Aschauer \emph{et al.} in CaMnO$_{3}$ thin films~\cite{aschauer13}.    

In view of the outcomes shown in Fig.~\ref{fig2}c, one might guess that, provided that the thermal 
contributions to the ${\rm V_{O}}$ free energy were more or less independent of epitaxial strain, 
tensile strain should favor the formation of vacancies at $T \neq 0$ conditions. However, as we 
have noted before, this is not what has been experimentally observed by Hu \emph{et al.} in 
SrCoO$_{3-\delta}$ thin films~\cite{hu16}. Fig.~\ref{fig3}a shows our vibrational entropy results
obtained for an equatorial vacancy, $F^{\rm qh}_{\rm vac}$, under compressive, neutral, and tensile 
strains expressed as a function of temperature; the following conclusions can be drawn from them. First, 
vibrational contributions to the ${\rm V_{O}}$ free energy strongly depend on the epitaxial strain; 
in the particular case of SCO those are most favorable in compressive thin films, contrarily to 
what is found for $E_{\rm vac}$. And second, at $T \approx 300$~K the $F^{\rm qh}_{\rm vac}$ 
difference between various $\eta$ states can be of the same order of magnitude, in absolute value, than 
the corresponding $E_{\rm vac}$ difference calculated at $T = 0$. For example, at room temperature 
the vibrational vacancy entropy estimated at $\eta = -2$\% is $16$ and $49$~meV/f.u. smaller than
the obtained in the unstrained and $+2$\% cases, respectively. These findings imply that as temperature 
is steadily increased the lattice excitations may entirely reverse the vacancy formation trends deduced 
at zero temperature. This is explicitly shown in Fig.~\ref{fig3}c, where at the highest temperature 
considered is easiest to create oxygen vacancies in compressively strained thin films (i.e.,  
$G^{* \rm qh}_{\rm vac}$ is minimum).         

In Fig.~\ref{fig3}b, we enclose the chemical expansion results obtained in SCO thin films. It 
is observed that $\Delta V_{\rm vac}$ is largest at compressive strains and that a small kink 
appears at the point in which the stoichiometric system undergoes a magnetic phase transition. 
As the vibrational properties of crystals largely depend on volume [i.e., the $V$-dependence 
of phonon frequencies, $\omega_{\boldsymbol{q}s}$, normally can be described with the Gr\"{u}neisen 
parameter $\gamma_{\boldsymbol{q}s} = -\partial(\ln \omega_{\boldsymbol{q}s}) / \partial(\ln V)$], we 
can reasonably correlate the observed $F^{\rm qh}_{\rm vac}$ strain dependence to the accompanying 
chemical expansion: the larger $\Delta V_{\rm vac}$ is, the more favorable lattice contributions
to ${\rm V_{O}}$ formation result. We comment again on this point below.    

\begin{figure}[t]
\centerline{
\includegraphics[width=1.0\linewidth]{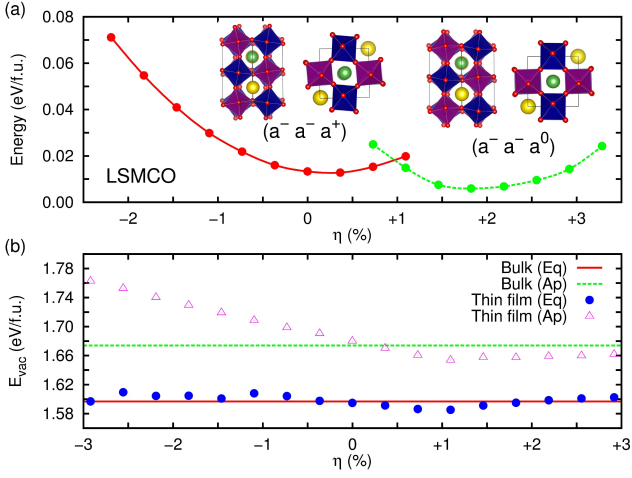}}
\caption{(Color online)~Zero-temperature calculations done in perfect and
         nonstoichiometric ($\delta = 0.25$) LSMCO thin films. Total 
         energy results obtained in stoichiometric systems expressed as a 
         function of epitaxial strain are shown in (a). The zero-temperature 
         vacancy formation energy is enclosed in (b). Both equatorial (Eq) and 
         apical (Ap) oxygen vacancy positions are considered and analogous 
         results obtained in bulk are shown for comparison.}
\label{fig4}
\end{figure}

\begin{figure}[t]
\centerline{
\includegraphics[width=1.0\linewidth]{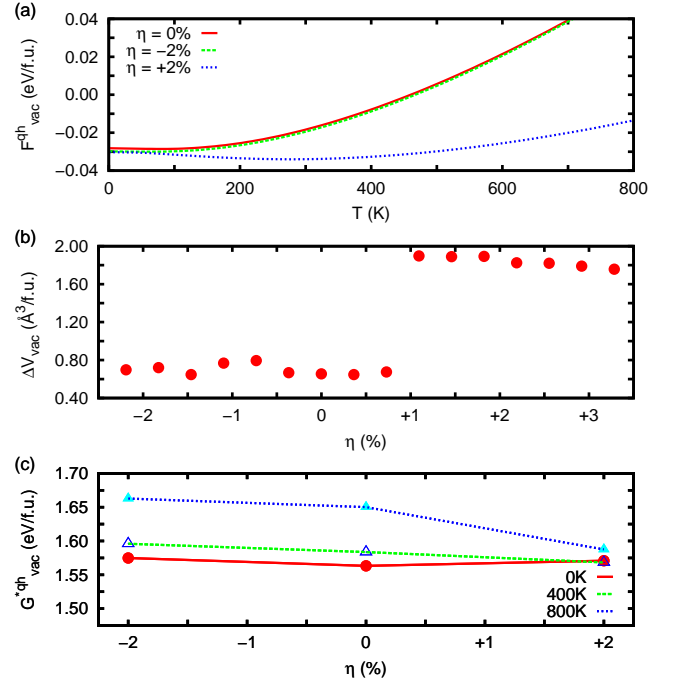}}
\caption{(Color online)~Quasi-harmonic free energy calculations performed in perfect and
         nonstoichiometric ($\delta = 0.25$, equatorial vacancies) LSMCO thin 
         films. Vibrational and thermodynamically shifted Gibbs vacancy free 
         energy results expressed as a function of temperature and epitaxial 
         strain are enclosed in (a) and (c), respectively. The accompanying 
         chemical expansion, $\Delta V_{\rm vac} = V_{\rm nonstoi} - V_{\rm perf}$, 
         expressed as a function of epitaxial strain is shown in (b).}
\label{fig5}
\end{figure}

{\bf La$_{0.5}$Sr$_{0.5}$Mn$_{0.5}$Co$_{0.5}$O$_{3-\delta}$ thin films}--Our DFT calculations 
predict an orthorhombic $Pbnm$ ground state for bulk stoichiometric LSMCO that is characterised 
by FM spin ordering and oxygen octahedral rotations $(a^{-}a^{-}a^{+})$, as expressed in Glazer's 
notation~\cite{glazer72} (i.e., antiphase within the substrate plane and in-phase in the 
perpendicular direction). At a tensile epitaxial strain of $\sim +1$\% the system undergoes a 
first-order transition to a monoclinic (and centrosymmetric) $P2/m$ phase, characterised also by 
FM spin ordering and O$_{6}$ rotations $(a^{-}a^{-}a^{0})$ [i.e., octahedral tiltings along the  
direction perpendicular to the substrate disappear, Fig.~\ref{fig4}a]. In the presence of oxygen 
vacancies the magnetic properties of LSMCO thin films remain invariant. With regard to $E_{\rm vac}$, 
we also find that (at $T = 0$) equatorial vacancies are created more easily than apical vacancies 
[i.e., $E_{\rm vac}({\rm Ap})-E_{\rm vac}({\rm Eq}) \sim 10$~meV/f.u.] although in this case the 
corresponding formation energy is quite insensitive to the strain conditions (Fig.~\ref{fig4}b).

Fig.~\ref{fig5}a shows our vibrational entropy results obtained for an equatorial vacancy, 
$F^{\rm qh}_{\rm vac}$, under compressive, neutral, and tensile strains expressed as a function 
of temperature. It is observed that lattice contributions to ${\rm V_{O}}$ formation now are 
practically identical in the unstrained and $\eta = -2$\% cases, and that $F^{\rm qh}_{\rm vac}$ 
is smallest for tensile strains. This behavior suggests that at finite temperatures equatorial 
vacancies in LSMCO thin films will be created more easily at $\eta > 0$ conditions, as it is 
explicitly shown in Fig.~\ref{fig5}c, due essentially to lattice effects. In this case we also 
find that the chemical expansion of the crystal (Fig.~\ref{fig5}b) and vibrational vacancy entropy 
are strongly correlated. In particular, the $\Delta V_{\rm vac}$ values computed in the unstrained 
and $\eta = -2$\% cases are almost identical and so are the corresponding $F^{\rm qh}_{\rm vac}$ 
curves. Meanwhile, the chemical expansion is larger at tensile strains, due to the underlying 
$Pbnm \to P2/m$ phase transition occurring at $\eta \sim +1$\%, and simultaneously the vibrational 
vacancy entropy becomes more favorable. Our \emph{ab initio} results obtained in SCO and LSMCO thin 
films clearly evidence a direct link between quantities $\Delta V_{\rm vac}$ and $F^{\rm qh}_{\rm vac}$.

Finally, it has been recently suggested, based on the results of zero-temperature $E_{\rm vac}$ 
calculations, that epitaxial strain could be used as a means to engineer vacancy ordering in perovskite 
oxides~\cite{aschauer13}. In fact, this appears to be the logical conclusion coming up from Figs.~\ref{fig2}c 
and~\ref{fig4}b. However, as we have already shown, thermal excitations can play a critical role on ${\rm V_{O}}$ 
formation at finite temperatures. In order to quantify the effects of lattice vibrations on 
possible vacancy ordering, we calculate the vibrational entropy of an apical (Ap) vacancy in 
SCO thin films at different strain states (Supplemental Material~\cite{supp}). Our $F^{\rm qh}_{\rm vac}$ 
results show that (i)~the vibrational entropy of Ap vacancies follows the same strain dependence 
than Eq vacancies, and importantly (ii)~lattice excitations strongly favor the creation of Ap 
vacancies over Eq; specifically, at medium and high temperatures we find that 
$F^{\rm qh}_{\rm vac}({\rm Ap})-F^{\rm qh}_{\rm vac}({\rm Eq}) \sim -10$~meV/f.u., which in absolute 
value is of the same order of magnitude than the corresponding static difference 
$E_{\rm vac}({\rm Ap})-E_{\rm vac}({\rm Eq})$. For instance, at $\eta = -2$\% and $T = 600$~K the 
vibrational entropy of Ap vacancies is about $25$~meV/f.u. smaller than that of Eq, and consequently 
$G^{* \rm qh}_{\rm vac}({\rm Ap}) \approx G^{* \rm qh}_{\rm vac}({\rm Eq})$. Therefore, lattice 
vibrations can hinder vacancy ordering (analogous results are found also in LSMCO thin films, 
Supplemental Material~\cite{supp}). We note that this conclusion remains valid when configurational 
vacancy contributions to $G^{* \rm qh}_{\rm vac}$ are taken into account, which always favor the 
formation of Eq vacancies over Ap, as those amount to only few meV/f.u. in the temperature interval 
of interest (Supplemental Material~\cite{supp}).

In summary, we have shown that lattice vibrations play a critical role on accurate prediction
of ${\rm V_{O}}$ formation in perovskite thin films. The variation of the vibrational vacancy 
entropy on epitaxial strain appears to be system dependent and can be strongly influenced by
underlying phase transitions. Nevertheless, we have revealed a direct link between $F^{\rm qh}_{\rm vac}$
and the chemical expansion of the system, $\Delta V_{\rm vac}$, that can be straightforwardly 
exploited in zero-temperature calculations to qualitatively foresee the effects of thermal 
excitations. In the two perovskite oxides analysed in this work, vibrational lattice excitations 
oppose to vacancy ordering.

The author wishes to thank N. A. Katcho for stimulating discussions. This research was 
supported by the Australian Research Council under Future Fellowship funding scheme 
(Grant No. FT140100135). Computational resources and technical assistance were provided 
by the Australian Government and the Government of Western Australia through Magnus under 
the National Computational Merit Allocation Scheme and The Pawsey Supercomputing Centre.

\end{document}